\documentclass{IEEEtran}
\usepackage[english]{babel}
\usepackage[utf8]{inputenc}
\usepackage{amsmath,amsfonts,amssymb,framed,graphicx,mathrsfs,mathtools,graphicx,multicol,blindtext,float,etoolbox,enumitem,gensymb, tikz, fancyhdr, graphicx, xfrac, multirow, url,blindtext,enumitem, listings,graphicx,tcolorbox,csquotes}
\usetikzlibrary{positioning}
\usepackage{color}
\definecolor{mygrey}{gray}{0.50}

\usepackage[braket,qm]{qcircuit}
\numberwithin{equation}{section}
\usepackage[hidelinks]{hyperref}
\usepackage[nameinlink, capitalise]{cleveref}
\usepackage[section]{placeins}

\usepackage[bibencoding=auto,maxbibnames=6,backend=biber,giveninits=true,sorting=none,doi=false,isbn=false,url=false]{biblatex}
\newcommand{\vt}[1]{\ensuremath{\boldsymbol{#1}}} % vector in juiste lettertype

\usepackage[T1]{fontenc}
\usepackage{Alegreya} %% Option 'black' gives heavier bold face 

\begin{document}
\onecolumn
\title{Distributed Quantum Machine Learning}
\author{\IEEEauthorblockN{Niels M. P. Neumann and Robert S. Wezeman}\\
\IEEEauthorblockA{\textit{Department of Applied Cryptography and Quantum Applications} \\
\textit{The Netherlands Organisation for Applied Scientific Research}\\
Anna van Buerenplein 1, 2595DA, The Hague, The Netherlands\\ 
\{niels.neumann, robert.wezeman\}@tno.nl
}}

\maketitle

\begin{abstract}
Quantum computers can solve specific complex tasks for which no reasonable-time classical algorithm is known. 
Quantum computers do however also offer inherent security of data, as measurements destroy quantum states. 
Using shared entangled states, multiple parties can collaborate and securely compute quantum algorithms. 
In this paper we propose an approach for distributed quantum machine learning, which allows multiple parties to securely perform computations, without having to reveal their data. 
We will consider a distributed adder and a distributed distance-based classifier. 
\end{abstract}

\begin{IEEEkeywords}
Distributed quantum computing, Quantum multi-party computation, Quantum machine learning, Quantum arithmetic
\end{IEEEkeywords}

\section{Introduction}\label{sec:introduction}
Quantum computers are in rapid development and the first classically intractable problems are already solved using quantum computers~\cite{Google_QuantumSupremacy_2019}. Even though these problems are artificial, specifically designed to show the power of quantum computers, it is expected that in the next few years, similar results will be achieved for practical problems. 

Apart from quantum computers, also in quantum internet there are rapid developments and the first small-scale networks are already realized~\cite{PHB_RealizationMultinodeQuantumNetwork_2021}. A quantum network allows for many new applications, including new forms of encryption~\cite{CGMO_PositionBasedCryptography_2009} and enhanced clock synchronization~\cite{JADW_QuantumClockSynchronization_2000,Chuang_QuantumDistributedClockSynchronization_2000}.

Quantum networks also allow for another application: distributed quantum computing, where different quantum computers are linked via a quantum network. We typically identify two types of distributed quantum computing. In the first, a single algorithm which is too large to be run on a quantum device, is subdivided in smaller parts, each of which can be run on a quantum device. In the second, multiple parties have access to local quantum computers which are linked via a quantum network. The parties can collaboratively perform quantum computations on their inputs without having to explicitly share it. 

The first type is a resource-problem. As hardware develops, larger problems can be run and the need to distribute the algorithms vanishes. The second type is the more interesting one as it opens up the way to completely new applications. In this work we will therefore focus on the second type of distributed quantum computing. 

Distributed quantum computing naturally extends classical multi-party computation, which allows multiple parties to collaborate securely~\cite{Yao_ProtocolsSecureComputations_1982}. We consider two applications of distributed quantum computing. The first being distributed arithmetic, the second being distributed distance-based classification. We show how both approaches work in a distributed setting and argue why information remains secure during the protocol execution. For both applications, multiple parties provide input and together perform the algorithm in such a way that the output is only revealed to one specific party without leaking information on individual parties input. 

In the next section, we give a brief introduction to quantum computing and some basic concepts of distributed quantum computing. In Section~\ref{sec:distributed_arithmetic} and Section~\ref{sec:distributed_classifier} we discuss a distributed quantum adder and a distributed distance-based classifier, respectively. In Section~\ref{sec:results} we provide a resource count of the distributed approaches. We conclude with some conclusions and an outlook. 

\section{Methods}\label{sec:methods}
\subsection{Brief Introduction to Quantum Computing}
Classical computers and quantum computers work similarly: Both operate on elementary units of computation and by performing the correct operations in the right order, problems can be solved. Classical computers operate on bits, zeros and ones, using classical gates, such as AND, OR and NOT gates. Quantum computers operate on qubits, superpositions of zeros and ones, using quantum operations, such as single qubit rotations and controlled-NOT (CNOT) gates. The CNOT gate is a two-qubit gate that flips the state of the second qubit, if the first qubit is in the one state.

A key difference between classical and quantum computers is that qubits do not have to be in one definite computational basis state $\ket{0}$ or $\ket{1}$. A qubit $\ket{\psi}$ can be in a superposition, a complex linear combination, of these basis states, which we can write as ${\ket{\psi} = \alpha\ket{0}+\beta\ket{1}}$ for $\alpha,\beta\in\mathbb{C}$ with $|\alpha|^2+|\beta|^2=1$. The computational basis states for multi-qubit states are given by ${\ket{x}=\ket{x_{n-1}}\otimes\hdots\otimes\ket{x_0}}$ for ${x_i\in\{0,1\}}$. Upon observing a quantum state, only one of the definite states is found. The probability to observe a given state equals the sum of the squared of the corresponding amplitudes. 

Another aspect in which quantum computers differ from classical ones is entanglement. Two or more quantum states can be correlated beyond what is possible classically. One of the most well-known entangled states is the GHZ-state given by
\begin{equation}
    \frac{1}{\sqrt{2}}(\ket{00}+\ket{11}),
\end{equation}
which is obtained by bringing the first qubit in a uniform superposition and then applying a CNOT gate. Upon measuring one qubit, the state of the other is instantaneously known, even if the entangled qubits are physically far apart. The GHZ-state naturally extends to more than two qubits and it has many applications. We will make use of GHZ-states to distribute gate operations to different quantum computers. 

We refer to~\cite{nielsen_chuang_2010} for a more elaborate introduction to quantum computing. 

\subsection{Distributed Quantum Computing\label{sec:DQC}}
To distributed quantum operations over multiple devices, it might seem natural to physically transport the qubits. However, as quantum states are fragile, this is likely to introduce errors. Instead, we propose to use shared entangled states. Two notable examples are quantum teleportation~\cite{QuantumTeleportation_1993} and non-local CNOT gates~\cite{Eisert:2000,YimsiriwattanaLomonaco:2004}. The quantum circuits for both approaches are similar and both use shared entangled states, for simplicity often assumed to be GHZ-states. 

The shared entangled states can also introduce noise in the computations. However, as they do not hold any information themselves, we can use techniques such as entanglement purification to minimize the effect of imperfect shared entanglement~\cite{BBPSSW_PurificationNoisyEntanglement_1996,BBPS_ConcentratingPartialEntanglement_1996,BDSW_MixedStateEntanglementPurification_1996}. These protocols focus the entanglement of multiple partially entangled states in fewer entangled states of higher quality using only local operations. We can generate these entangled state of high quality before we start a distributed algorithm, and hence the effect of noise on the data is limited. 

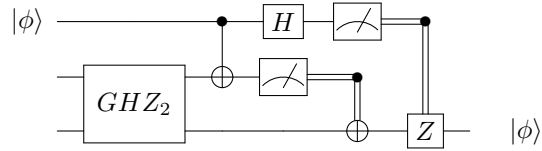
\begin{figure}[t!]
	\centering
	$${\Qcircuit @C=1em @R=0.7em {
	    \lstick{\ket{\phi}} & \qw & \ctrl{1} & \gate{H} & \meter & \control\cw\cwx[2] \\
		& \ghost{GHZ_2} & \targ & \meter & \control\cw\cwx[1] &&&& \\
		& \multigate{-1}{GHZ_2} & \qw & \qw & \targ & \gate{Z} & \qw & \rstick{\ket{\phi}}
	}}$$
	\caption{A quantum circuit that teleports a state $\ket{\phi}$ to a third qubit. Double lines indicate classical information.}
	\label{fig:circuitTeleportation}
\end{figure}
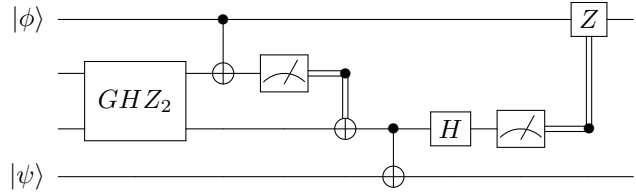
\begin{figure}[t!]
	\centering
	${\Qcircuit @C=1em @R=0.7em {
	    \lstick{\ket{\phi}} & \qw & \ctrl{1} & \qw & \qw & \qw & \qw & \qw & \gate{Z} & \qw & \\
		& \ghost{GHZ_2} & \targ & \meter & \control\cw\cwx[1] &&&& \\
		& \multigate{-1}{GHZ_2} & \qw & \qw & \targ & \ctrl{1} & \gate{H} & \meter & \control\cw\cwx[-2] \\
		\lstick{\ket{\psi}} & \qw & \qw & \qw & \qw & \targ & \qw & \qw & \qw & \qw & 
	}}$
	\caption{A quantum circuit for the non-local CNOT-gate between $\ket{\phi}$ and $\ket{\psi}$. The top and bottom qubit might belong to different quantum devices that share a GHZ-state. Double lines represent classical information.}
	\label{fig:circuitNonLocalCNOT}
\end{figure}

The quantum circuits for quantum teleportation and the non-local CNOT gate are shown in~\cref{fig:circuitTeleportation} and~\cref{fig:circuitNonLocalCNOT}, respectively. Both circuits extend naturally to more parties. The two circuits show similarities, however their effect is different. The first circuit teleports a quantum state from one qubit to another. These qubits can be hosted on different devices. The first qubit is measured and hence destroyed. Distributed computations can be performed by teleporting the quantum state from one device to another and then perform all operations locally on the second device. 
The second circuit applies a two-qubit CNOT operation between one qubit and another that can, again, be hosted on another device. Both qubits remain coherent and further operatoins can be applied to them. 

It is known that single qubit rotations and a multi-qubit gate, such as the CNOT gate are universal for quantum computing~\cite{Barenco:1995}. This means that any quantum operation can be broken down in a sequence of single qubit gates and two-qubit gates. 
We can hence distribute quantum algorithms once we can distribute the two-qubit operations. In this work, we used non-local CNOT gates to implement distributed algorithms. This way, the qubit remains intact on the first device and we do not require a second step of teleportation to get the qubit back to the first device again. 

Note that a non-local CNOT gate does not create an independent copy of a state. Instead, it creates an entangled state, i.e., given a state $\ket{\phi}=\alpha\ket{0}+\beta\ket{1}$ the non-local CNOT gate implements the map
\begin{equation}
    \ket{\phi}\ket{0}\mapsto \alpha\ket{00}+\beta\ket{11}.
\end{equation}

A property that we will use later is that phases applied to any qubit of a GHZ-state have a global effect: A phase applied to the first qubit of a GHZ-state results in the same state as when that phase was applied to any of the other qubits of the GHZ-state. This is independent of the physical location of the qubits: The qubits might even be hosted on different devices. 

We will use this property together with the non-local CNOT-gate to distribute operations among various parties. in the next two sections, we will show how to do this for two quantum machine learning applications. 

\section{Distributed Adder\label{sec:distributed_arithmetic}}
We first consider how to perform arithmetics with multiple parties. This simple yet relevant topic provides a first insight in the value of distributed quantum computing. It is also a relevant topic, as arithmetics form an important pillar in many algorithms, for instance to compute the mean of a set of number~\cite{Kiltz:2005}. 

In this section we will solely focus on adding numbers. More complex arithmetics such as multiplication follow naturally by repeated addition. Given two basis states $\ket{a}$ and $\ket{b}$, a quantum adder implements ${\ket{a}\ket{b}\mapsto\ket{a}\ket{a+b}}$. The addition is modulo $2^N$ by definition, with $N$ the number of qubits per register. By linearity of quantum computing, a quantum adder also works on arbitrary superpositions. 

We can implement a quantum adder in multiple ways. Most approaches do however require gates that are not directly supported by underlying hardware. Draper~\cite{Draper:2000} presented an approach that only uses single qubit gates and controlled-phase gates. Later this work was extended to multiplication and modular arithmetic~\cite{Draper:2000,Beauregard:2003,RuizPerez:2017}. This applies a quantum Fourier transform on a quantum state $\ket{b}$ to obtain $\ket{\phi(b)}$. On this transformed state, we can now apply controlled phase-gates with predetermined phases, targets and controls, to add integers. An inverse quantum Fourier transform then gives the desired quantum state. 

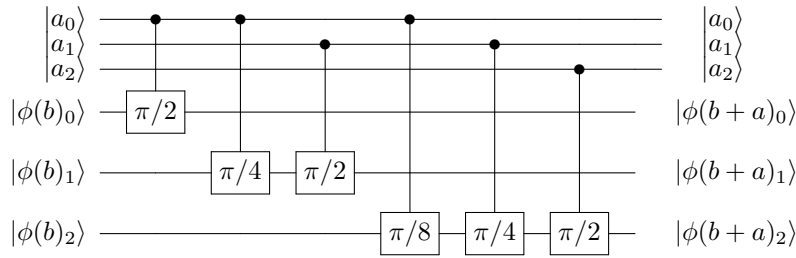
\begin{figure}[th]
	\centering
	\begin{minipage}[c]{0.5\textwidth}
		\centering
		{\Qcircuit @C=1em @R=0.7em {
			\lstick{\ket{a_0}} & \ctrl{3} & \ctrl{4} & \qw & \ctrl{5} & \qw & \qw & \qw & \qw & \rstick{\ket{a_0}} \\
			\lstick{\ket{a_1}} & \qw & \qw & \ctrl{3} & \qw & \ctrl{4} & \qw & \qw & \qw & \rstick{\ket{a_1}} \\
			\lstick{\ket{a_2}} & \qw & \qw & \qw & \qw & \qw & \ctrl{3} & \qw & \qw & \rstick{\ket{a_2}} \\
			\lstick{\ket{\phi(b)_0}} & \gate{\pi/2} & \qw & \qw & \qw & \qw & \qw & \qw & \rstick{\ket{\phi(b+a)_0}} \\
			\lstick{\ket{\phi(b)_1}} & \qw & \gate{\pi/4} & \gate{\pi/2} & \qw & \qw & \qw & \qw & \rstick{\ket{\phi(b+a)_1}} \\
			\lstick{\ket{\phi(b)_2}} & \qw & \qw & \qw & \gate{\pi/8} & \gate{\pi/4} & \gate{\pi/2} & \qw & \rstick{\ket{\phi(b+a)_2}}
		}}
	\end{minipage}
	\caption{The addition part of a quantum Fourier transform-based adder. $\ket{\phi(b)_j}$ represents the $j$-th bit of the Fourier transform of $b$. The gates represent controlled-$R_Z$ gates, where the argument shown is the argument of the $R_Z$-gate. A quantum Fourier transform is needed to retrieve the quantum state $\ket{a}\ket{b+a}$.}
	\label{fig:addition_fourier_state}
\end{figure}

The quantum circuit for addition is shown in~\cref{fig:addition_fourier_state}. The blocks represent $R_Z$-gates, with the argument shown inside the block. The matrix representation of these gates is
\begin{equation}
    R_Z(\theta) = \begin{pmatrix} 1 & 0 \\ 0 & e^{2i\theta} \end{pmatrix}.
\end{equation}
As we use controlled operations, they are only applied if the controlling qubit is in the $\ket{1}$ state.
The state $\ket{\phi(b)_j}$ represents the $j$-th qubit of the quantum Fourier transformed state $\ket{\phi(b)}$, given by $\ket{\phi(b)_j} = \frac{1}{\sqrt{2}}\left(\ket{0} + \exp({2\pi i \cdot b / 2^{j+1}}) \ket{1}\right)$. 
Note that for the $j$-th qubit of the quantum Fourier transformed state, only the first $j$ bits of $a$ and $b$ are relevant. 
All other bits will contribute an integer value to the fraction $b/2^j$ and hence will have no physical effect. 
After the phase-gates are applied, the state of the $j$-th qubit is given by
\begin{equation}
    \frac{1}{\sqrt{2}}\left(\ket{0} + e^{\pi i \cdot (b+a) / 2^j} \ket{1}\right) = \ket{\phi(b+a)_j}.
\end{equation}
An inverse quantum Fourier transform indeed gives the $j$-th bit of the sum $b+a$. 

Even though the shown circuit only concerns computational basis states, it works equally well for superpositions as input. For our application, computational basis states are sufficient and this also allows us to apply the phase-gates controlled by classical bits instead of by qubits. This method extends naturally to adding multiple numbers in parallel by using larger GHZ-states and distributing the quantum state to more parties.

Suppose we have $K$ parties, each with $N$ qubits. Additionally, we have one server party that prepares and distributes the initial quantum state using GHZ-states and non-local CNOT gates. After distribution, each party applies the phases to their part of the shared entangled state. The server party then again uses GHZ-states and non-local CNOT gates to assure that an inverse quantum Fourier transform indeed gives the correct sum. The distributing operation of the server party thus implements the following map for each of the $N$ bits:
\begin{equation}
\ket{\phi(b)_j}\ket{0}^{\otimes K} \mapsto \frac{1}{\sqrt{2}}\left(\ket{0}^{\otimes K+1}+ e^{\pi i \cdot b / 2^j} \ket{1}^{\otimes K+1}\right).
\label{eq:fan_out}
\end{equation}

After each party has locally applied the phase gates corresponding to their input, we are left with the quantum state
\begin{equation}
\frac{1}{\sqrt{2}}\left(\ket{0}^{\otimes K+1}+ e^{\pi i \cdot (b + \sum_k x^k) / 2^j} \ket{1}^{\otimes K+1}\right),
\label{eq:after_phase_gate_round}
\end{equation}
where $x^k$ is the input of party $k$. The server party uses GHZ-states and non-local CNOT gates to obtain
\begin{equation}
\frac{1}{\sqrt{2}}\left(\ket{0}+ e^{\pi i \cdot (b + \sum_k x^k) / 2^j} \ket{1}\right)\ket{0}^{\otimes K}.
\label{eq:fan_out_back}
\end{equation}
An inverse quantum Fourier transform now indeed gives the sum $b+\sum_k x^k$, as desired. 

If we would omit this operation, the remaining entanglement would stop states from cancelling out under the inverse quantum Fourier transform and we would be left with the wrong final state. 

The server can follow two approaches to distribute the operations and allow multiple parties to add their input. In the first approach, shown in~\cref{fig:addition:method_1}, non-local CNOT gates with more targets are used. In the second approach, shown in~\cref{fig:addition:method_2}, more non-local CNOT gates with only one target are used. 
\begin{figure}[th]
	\centering
	\begin{minipage}[c]{0.5\textwidth}
		\centering
		{\Qcircuit @C=0.5em @R=0.45em {
			\lstick{s:\ket{0}} & \multigate{2}{QFT} & \ctrl{3} & \qw & \qw & \qw & \ctrl{3} & \qw & \qw & \multigate{2}{QFT^{-1}} & \qw & \\
			\lstick{s:\ket{0}} & \ghost{QFT} & \qw & \ctrl{3} & \qw & \qw & \qw & \ctrl{3} & \qw & \ghost{QFT^{-1}} & \qw & \\
			\lstick{s:\ket{0}} & \ghost{QFT} & \qw & \qw & \ctrl{3} & \qw & \qw & \qw & \ctrl{3} & \ghost{QFT^{-1}} & \qw & \\
			\lstick{p_1:\ket{0}} & \qw & \targ\qwx[3] & \qw & \qw & \multigate{2}{\text{Add } x^1} & \targ\qwx[3] & \qw & \qw & \qw & \qw & \rstick{\ket{0}} \\
			\lstick{p_1:\ket{0}} & \qw & \qw & \targ\qwx[3] & \qw & \ghost{\text{Add } x^1} & \qw & \targ\qwx[3] & \qw & \qw & \qw & \rstick{\ket{0}} \\
			\lstick{p_1:\ket{0}} & \qw & \qw & \qw & \targ\qwx[3] & \ghost{\text{Add } x^1} & \qw & \qw & \targ\qwx[3] & \qw & \qw & \rstick{\ket{0}} \\
			\lstick{p_2:\ket{0}} & \qw & \targ & \qw & \qw & \multigate{2}{\text{Add } x^2} & \targ & \qw & \qw & \qw & \qw & \rstick{\ket{0}} \\
			\lstick{p_2:\ket{0}} & \qw & \qw & \targ & \qw & \ghost{\text{Add } x^2} & \qw & \targ & \qw & \qw & \qw & \rstick{\ket{0}} \\
			\lstick{p_2:\ket{0}} & \qw & \qw & \qw & \targ & \ghost{\text{Add } x^2} & \qw & \qw & \targ & \qw & \qw & \rstick{\ket{0}} \\
		}}
	\end{minipage}
	\caption{Addition of integers by two parties, using one server party. We use CNOT gates with multiple targets. The blocks $\text{Add }x^k$ stand for the phase-gates needed to add integer $x^k$, similar to~\cref{fig:addition_fourier_state}. The final quantum state in the first register is $\ket{{x^1 + x^2}}$.}
	\label{fig:addition:method_1}
\end{figure}
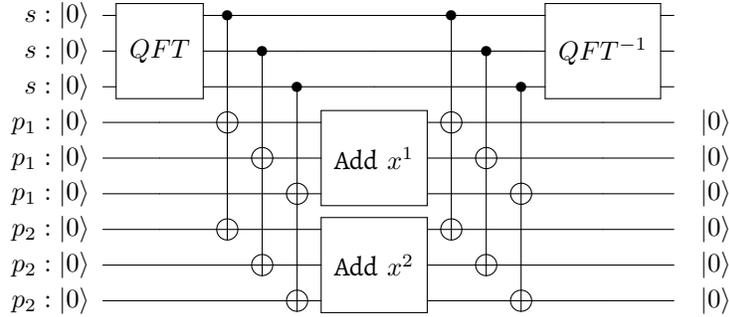
\begin{figure}[th]
	\centering
	\begin{minipage}[c]{0.5\textwidth}
		\centering
		{\Qcircuit @C=0.4em @R=0.45em {
			\lstick{s:\ket{0}} & \multigate{2}{QFT} & \ctrl{3} & \qw & \qw & \qw & \ctrl{3} & \qw & \qw & \ctrl{6} & \qw & \qw & \qw & \ctrl{6} & \qw & \qw & \multigate{2}{QFT^{-1}} & \qw & \\
			\lstick{s:\ket{0}} & \ghost{QFT} & \qw & \ctrl{3} & \qw & \qw & \qw & \ctrl{3} & \qw & \qw & \ctrl{6} & \qw & \qw & \qw & \ctrl{6} & \qw & \ghost{QFT^{-1}} & \qw & \\
			\lstick{s:\ket{0}} & \ghost{QFT} & \qw & \qw & \ctrl{3} & \qw & \qw & \qw & \ctrl{3} & \qw & \qw & \ctrl{6} & \qw & \qw & \qw & \ctrl{6} & \ghost{QFT^{-1}} & \qw & \\
			\lstick{p_1:\ket{0}} & \qw & \targ & \qw & \qw & \multigate{2}{\text{Add } x^1} & \targ & \qw & \qw & \qw & \qw & \qw & \qw & \qw & \qw & \qw & \qw & \qw & \rstick{\ket{0}} \\
			\lstick{p_1:\ket{0}} & \qw & \qw & \targ & \qw & \ghost{\text{Add } x^1} & \qw & \targ & \qw & \qw & \qw & \qw & \qw & \qw & \qw & \qw & \qw & \qw & \rstick{\ket{0}} \\
			\lstick{p_1:\ket{0}} & \qw & \qw & \qw & \targ & \ghost{\text{Add } x^1} & \qw & \qw & \targ & \qw & \qw & \qw & \qw & \qw & \qw & \qw & \qw & \qw & \rstick{\ket{0}} \\
			\lstick{p_2:\ket{0}} & \qw & \qw & \qw & \qw & \qw & \qw & \qw & \qw & \targ & \qw & \qw & \multigate{2}{\text{Add } x^2} & \targ & \qw & \qw & \qw & \qw & \rstick{\ket{0}} \\
			\lstick{p_2:\ket{0}} & \qw & \qw & \qw & \qw & \qw & \qw & \qw & \qw & \qw & \targ & \qw & \ghost{\text{Add } x^2} & \qw & \targ & \qw & \qw & \qw & \rstick{\ket{0}} \\
			\lstick{p_2:\ket{0}} & \qw & \qw & \qw & \qw & \qw & \qw & \qw & \qw & \qw & \qw & \targ & \ghost{\text{Add } x^2} & \qw & \qw & \targ & \qw & \qw & \rstick{\ket{0}} \\
		}}
	\end{minipage}
	\caption{Addition of integers by two parties, using one server party. We use CNOT gates with a single target and as such require the parties to add their input sequentially. The blocks $\text{Add }x^k$ stand for the phase-gates needed to add integer $x^k$, similar to~\cref{fig:addition_fourier_state}. The final quantum state in the first register is $\ket{{x^1 + x^2}}$.}
	\label{fig:addition:method_2}
\end{figure}
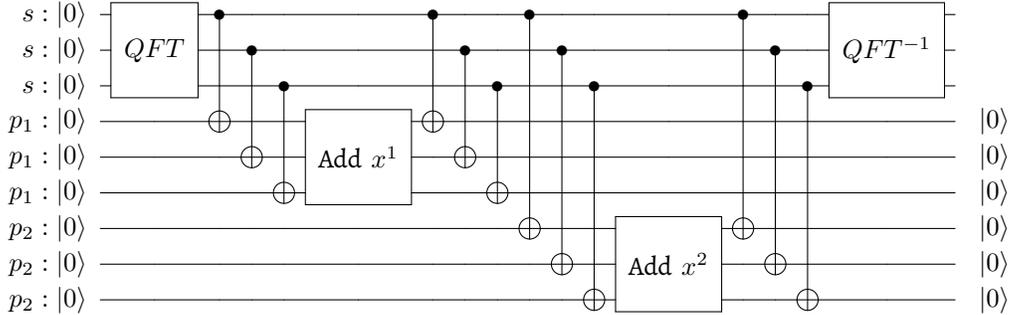

The difference between both approaches becomes clear when we replace the CNOT-gates by non-local variants. In the first approach shown in~\cref{fig:addition:method_1}, we require larger GHZ-states, but less in total. In the second approach shown in~\cref{fig:addition:method_2}, we require smaller GHZ-states, which are easier to create, however, we require more of them.

Note that in the shown circuits we used a single designated server party to create and distribute the initial state. It is also this server party that learns the sum of the inputs, whereas the other players will measure only the zero-state. We can allow the server party to also provide input for the summation, which, in some use-cases, might be practical. The analysis remains the same. 

None of the parties can learn the input of other parties, including the server party. This follows as the information is encoded in the phases of the quantum states. Only after we have the state of~\eqref{eq:fan_out_back} and apply an inverse quantum Fourier transformation on it, can we access the sum of the inputs. Otherwise, measurements will return no useful information. 

Finally, note that only the server can have a non-zero measurement outcome if the procedure is performed correctly. The server party learns only the sum of all inputs, but learns no information on the input of individual parties. 

\section{Distributed Distance-Based Classifier}\label{sec:distributed_classifier}
The second distributed machine learning approach is a distributed classifier. We consider a distance-based classifier~\cite{SFP_DistanceBasedClassifier_2017}. Given $N$ normalised data points $\vt{x}_i$, each having $M$ features, and labels $y_i\in\{-1,1\}$, this classifier computes a label $\tilde{y}$ for a new data point $\tilde{\vt{x}}$ by evaluating the function
\begin{equation}\label{eq:sgn_kernel}
    \tilde{y} = \mathrm{sgn}\left( \sum_i y_i \Big[ 1-\frac{1}{4N}|\tilde{\vt{x}} - \vt{x}_i|^2\Big] \right).
\end{equation}

We can evaluate this function and determine the new label by manipulating the initial quantum state
\begin{equation}\label{eq:initialState}
    \ket{\psi} = \frac{1}{\sqrt{2N}} \sum_{n=0}^{N-1} \ket{n}\Big(\ket{0}\ket{\psi_{\tilde{\vt{x}}}} + \ket{1}\ket{\psi_{\vt{x}_n}}\Big)\ket{y_n}.
\end{equation}
Here $\ket{\psi_{x}}$ is the amplitude encoding of a normalised data point $\vt{x}$: $\ket{\psi_{\vt{x}}} = \sum_i x_i\ket{i}$. This state requires $n=\log_2 N$ qubits for the data point counter and $m=\log_2 M$ qubits for the representation of a data point. In total $n+m+2$ qubits are required. To evaluate the kernel function, we first apply a Hadamard gate on the second register and then measure the second and fourth register. If the measurement of the second register equals zero, the probability distribution of the measurement of the last register corresponds precisely to the sum of the magnitude of the positive and negative terms of~\eqref{eq:sgn_kernel}. 

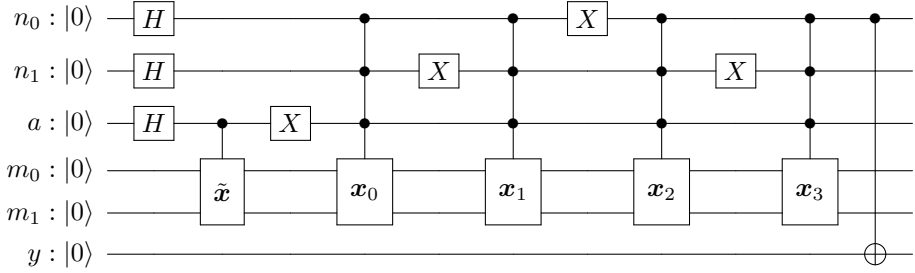
\begin{figure}
    \centering
    	\begin{minipage}[c]{0.5\textwidth}
		\centering
		{\Qcircuit @C=1em @R=0.7em{
			\lstick{n_0: \ket{0}} & \gate{H} & \qw & \qw & \ctrl{1} & \qw & \ctrl{1} & \gate{X} & \ctrl{1} & \qw & \ctrl{1} & \ctrl{5} & \qw\\
			\lstick{n_1: \ket{0}} & \gate{H} & \qw & \qw & \ctrl{1} & \gate{X} & \ctrl{1} & \qw & \ctrl{1} & \gate{X} & \ctrl{1}& \qw & \qw\\
			\lstick{a: \ket{0}}   & \gate{H} & \ctrl{1} & \gate{X} & \ctrl{1} & \qw & \ctrl{1} & \qw & \ctrl{1} & \qw & \ctrl{1}& \qw & \qw\\
			\lstick{m_0: \ket{0}} & \qw      & \multigate{1}{\tilde{\vt{x}}} & \qw & \multigate{1}{\vt{x}_0} & \qw & \multigate{1}{\vt{x}_1} & \qw & \multigate{1}{\vt{x}_2} & \qw & \multigate{1}{\vt{x}_3}& \qw & \qw\\
			\lstick{m_1: \ket{0}} & \qw      & \ghost{\vt{x}} & \qw & \ghost{\vt{x}_0} & \qw & \ghost{\vt{x}_1} &\qw & \ghost{\vt{x}_2} & \qw & \ghost{\vt{x}_3} & \qw  & \qw\\
			\lstick{y: \ket{0}}   & \qw      & \qw & \qw & \qw & \qw & \qw & \qw & \qw & \qw & \qw & \targ{} & \qw\\ 
		}}
	\end{minipage}
    \caption{Initial state preparation for the distance-based classifier for four data points, each having four features. Each (multi-)controlled block represents amplitude encoding of a normalised data point $\vt{x}$.}
    \label{fig:dbc}
\end{figure}

The power of this algorithm is especially clear if the initial quantum state is given as starting point. Otherwise we have to explicitly prepare it. Multiple extensions of this distance-based classifier have since been proposed~\cite{Wezeman:2019,BlankSilva:2022} extending the number of data points, the number of features and removing the need for a label qubit. An example how the state can be prepared is shown in ~\cref{fig:dbc} for the case of four data points each having four features.

\begin{figure}
    \centering
    \begin{minipage}[c]{0.8\textwidth}
    \Qcircuit @C=1em @R=.7em {
    & \gate{R_Y\left(\alpha_1\right)} & \ctrlo{1} & \ctrl{1} & \qw \\
    & \qw & \gate{R_Y\left(\alpha_{2,0}\right)} & \gate{R_Y\left(\alpha_{2,1}\right)} & \qw 
    }
    \end{minipage}
    \caption{Generic circuit for amplitude encoding of a normalised data point $x$ with four features. Using trigonometry the angles $\alpha$ can be calculated~\cite{Sun2001}.}
    \label{fig:encode}
\end{figure}
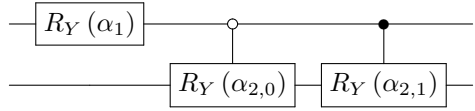

An important aspect of creating this initial state is the amplitude encoding of data points. Amplitude encoding of arbitrary data points can be obtained using only $R_Y$- and controlled $R_Y$-gates, as was shown by \cite{Sun2001}. An example of such a circuit for a data point with four features is shown in~\cref{fig:encode}.

% Describe distributed setting
We now consider the case where multiple parties wish to evaluate this distance-based classifier using their input data. The input data from these parties can either be horizontally or vertically separated. Horizontally separated data means that each party has some of the different data points, while for those data points having access to all features. Vertically separated data means that parties have information for the same data points but each party now has access to different features of that data point. We only consider the case of horizontally separated data, however, extensions to vertically separated data are possible. 

\begin{figure}
    \centering
    \begin{minipage}[c]{0.8\textwidth}
    \Qcircuit @C=0.5em @R=.7em {
    & \qw & \ctrl{1} & \qw & \qw & & & & & \qw & \qw  & \ctrl{1} & \qw & \ctrl{1} & \qw & \qw \\
    & \gate{R_x(\frac{\pi}{2})} & \gate{R_z(\alpha)} & \gate{R_x(-\frac{\pi}{2})} & \qw & & \lstick{\equiv}& & & \qw & \gate{R_x(\frac{\pi}{2})} & \ctrl{1} & \qw & \ctrl{1} & \gate{R_x(-\frac{\pi}{2})} & \qw \\
    \lstick{\ket{0}} & \qw & \qw & \qw & \qw & & & & \lstick{\ket{0}} & \qw & \qw & \targ{} & \gate{R_z(\alpha)} & \targ{} & \qw & \qw &   \\
    }
    \end{minipage}
    \caption{Circuit identity for a controlled-$R_Y$ rotation.}
    \label{fig:distributedRy}
\end{figure}
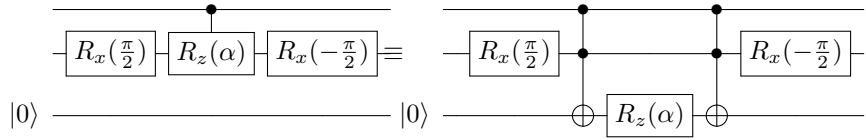

Consider the following setting: One computing server party wishes to classify a data point $\tilde{\vt{x}}$ based on input data that is provided by $K$ different parties. We assume that each of these $K$ parties provides $N$ data points each having $M$ features. Their goal remains to create the state of~\eqref{eq:initialState} on the device of the server party, where the only difference is that now the data encoding is performed in a distributed manner. From the amplitude encoding circuit we can see that this can be done if it is possible to perform controlled-$R_Y$ operations in a distributed manner. To do so, we use that rotations around the $Y$-axis can be replaced by rotations around the $Z$-axis given a suitable basis transformation: $R_Y(\theta) = R_X(-\frac{1}{2}\pi)R_Z(\theta)R_X(\frac{1}{2}\pi)$. A circuit for controlled $R_Y$-operation using $R_Z$ gates is shown in~\cref{fig:distributedRy}.

Suppose that the first two qubits belong to the device of the server party and the third qubit belongs to a device of one of the $K$ parties. We can distribute a controlled $R_Y$-operation using the shown circuit and by then replacing the controlled-gates by distributed controlled-gates, as introduced in Subsection~\ref{sec:DQC}. It is then straightforward to use this circuit as a building block to implement amplitude encoding in a distributed manner. Note that each party requires only their share of a GHZ-state and one additional qubit, independent of the number of features.

\begin{figure}
    \centering
		\includegraphics{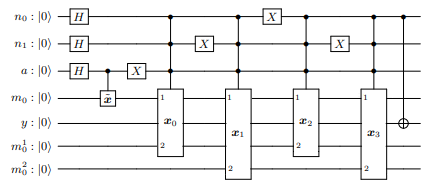}
    \caption{Initial state preparation for distributed distance-based classifier. Each party provides two data points each with two features. The first five qubits are hosted on the server, qubit six and seven are controlled by party one and party two respectively.}
    \label{fig:distributedDBC}
\end{figure}

An example of the final circuit used to obtain the initial state from two data providing parties is shown in~\cref{fig:distributedDBC}. In this figure we have not explicitly drawn the needed GHZ qubits needed to perform the distributed data encoding. Only the qubits that are numbered in the multi-qubit gates are involved in the encoding operation. 

\section{Results}\label{sec:results}
To show the potential of the distributed quantum adder from Section~\ref{sec:distributed_arithmetic} and the distributed distance-based classifier from Section~\ref{sec:distributed_classifier}, we implemented them and compared the results to those of a simulation with all computations performed on a single device. For the final result is should not matter whether computations are performed distributed or whether they are run on a single larger quantum computer. We simulated both distributed machine learning algorithms using Qiskit~\cite{Qiskit} and found that the final quantum states and the measurements thereof were identical for both the distributed versions and the local versions.

We considered multiple ways to distribute operations, for instance using bigger GHZ-states and applying a non-local CNOT gate with multiple targets at once, or using multiple 2-party GHZ-states and apply multiple smaller non-local CNOT gates sequentially. Below we consider the resource requirements for both approaches. 

\subsection{Distributed Adder}
For the distributed quantum adder, we consider the case with a server party and $K$ different parties, each inputting an integer. Let $N$ be the maximum bit size of integers. The total number of qubits that are needed to represent the output and input from the $K$ parties is $N\cdot (K+1)$ qubits. The circuit contains $2N \cdot K$ CNOT operations. We consider four different methods to implement the distributed circuit:
\begin{enumerate}
    \item Use local CNOT gates. This is not a distributed implementation and is used as benchmark case;
    \item Use 2-party GHZ-states and apply multiple sequential non-local CNOT gates with a single target. This method increases the total number of qubits by $4N\cdot K$;
    \item Use $K+1$-party GHZ-states and apply non-local CNOT gates with $K$ targets, one for each party. This method increases the total number of qubits by $2N\cdot(K+1)$;
    \item Use $K+1$-party GHZ-states and apply non-local CNOT gates with $K$ targets. Qubits used for the GHZ-states are reused. This method increases the total number of qubits by $K+1$. 
\end{enumerate}

The latter method is especially useful for simulation purposes. In practice it might be challenging to reuse qubits. The resource requirements for these different methods are summarized in~\cref{tab:distributedadder}.
\begin{table}
\centering
\caption{Required number of qubits and GHZ-states to implement the (distributed) quantum adder using different methods to distribute operations.}
\begin{tabular}{l|c|c}
Method & Total number of qubits & Number of GHZ states  \\ \hline
1 & $N\cdot (K+1)$    &           $0$                                         \\
2 & $N\cdot (5K+1)$   &           $2N\cdot K$ GHZ\textsubscript{2}            \\
3 & $N\cdot (3K+3)$   &           $2N\cdot N$ GHZ\textsubscript{K+1}          \\
4 & $(N+1)\cdot(K+1)$ &           $2N\cdot N$ GHZ\textsubscript{K+1}          \\
\end{tabular}
    \label{tab:distributedadder}
\end{table}

\subsection{Distributed Distance-Based Classifier}
For the distributed distance-based classifier, we consider the case where the initial state~\eqref{eq:initialState} has to be prepared explicitly on a server parties quantum device. We consider $K$ parties with each $N$ horizontally separated data points having $M$ features each. The server needs $\log_2 (K\cdot N) + 1 + \log_2(M) + 1$ qubits. The $K$ different parties only need one qubit each and the additional qubits for the GHZ states. The total number of (distributed) CNOT operations that are needed can be calculated by the number of data point that need to be encoded, $K\cdot N$, and multiply it by twice the number of $R_Y$ gates that are needed to encode a data point with $M$ features: $M-1$. 

Similar to the distributed adder, we consider two methods of implementing CNOT operations
\begin{enumerate}
    \item Use local CNOT gates. This is not a distributed implementation and is used as benchmark case;
    \item Use 2-party GHZ-states and apply multiple non-local CNOT gates with a single target. This method increases the total number of qubits with $2 N\cdot K \cdot (M-1)$;
    \item Use 2-party GHZ-states and apply multiple non-local CNOT gates with a single target. Qubits used for the GHZ-states are reused. This method increases the total number of qubits by $2$.
\end{enumerate}

The results for these different methods are summarized in~\cref{tab:distributedclassifier}.
\begin{table}
\centering
    \caption{Required number of qubits and GHZ-states to implement the (distributed) distance-based-classifier using different methods to distribute operations.}
\begin{tabular}{l|c|c}
Method & Total number of qubits & Number of GHZ states \\ \hline
1 & $\log_2 (K\cdot N) + \log_2(M) + 2$                         & $0$ \\
2 & $\log_2 (K\cdot N) + \log_2(M) + 2 + 2N\cdot K\cdot (M-1)$  & $N\cdot K \cdot (M-1)$ GHZ\textsubscript{2} \\
3 & $\log_2 (K\cdot N) + \log_2(M) + 4$                         & $N\cdot K \cdot (M-1)$ GHZ\textsubscript{2} \\
\end{tabular}
    \label{tab:distributedclassifier}
\end{table}

\section{Conclusion}
In this work we considered two applications of distributed quantum computing: distributed arithmetics and distributed distance-based classifiers. We showed how GHZ-states can allow multiple parties to simultaneously perform computations without having to directly share their data. Instead of physically having to share (encrypted) data, parties can apply local phase-gates and in that way share their data. These individual phases are immeasurable and only after all parties inputted their data and a suitable transformation is applied, will we learn the outcome. 

Our simulations showed that the distributed approaches indeed give the same answer as when the computations would be run locally. More research on the practical implementation of these methods is however needed, as then other aspects, including decoherence will play a more prominent role. 

A final important aspect of great importance is the security of the data encoded by the different parties. Each party adds a phase to their part of the entangled state. These phases are however immeasurable in its current form. Only after a suitable operation can we learn this information. Intermediate adverserial measurements destroy the quantum state and information encoded in the phases is lost. 

Our work showed the potential of distributed quantum machine learning. More interesting application exist, such as federated quantum machine learning. Another future research direction is incorporating physical effects in the simulations. We already briefly discussed this topic in Sectionion~\ref{sec:methods} on why shared entangled states are better than actually transporting data qubits. It is important to know the cost of creating shared entangled states of sufficient quality and their effect on the quality of the algorithm. 

\printbibliography
\end{document}